\documentclass[iop]{emulateapj}
\usepackage{natbib,amssymb,amsmath,graphicx}
\begin{document}

\title{Far Infrared Variability of Sagittarius A*: 25.5 hours of Monitoring
with \textit{Herschel}$^{\dagger}$}\thanks{$^{\dagger}${\it Herschel} is an ESA space observatory with
science instruments provided by European-led Principal Investigator consortia
and with important participation from NASA.}

\author{Jordan M. Stone, D.~P. Marrone} 
\affil{Steward Observatory,
University of Arizona,
933 N. Cherry Ave,
Tucson, AZ 85721-0065 USA;
jstone@as.arizona.edu, dmarrone@as.arizona.edu}

\author{C.~D. Dowell} 
\affil{Jet Propulsion Laboratory, 
4800 Oak Grove Drive, 
Pasadena, CA 91109, USA; 
charles.d.dowell@jpl.nasa.gov}

\author{B.Schulz}
\affil{Infrared Processing and Analysis Center, 
MS 100-22, 
California Institute of Technology, 
JPL, 
Pasadena, CA 91125, USA; bschulz@ipac.caltech.edu}

\author{C.~O. Heinke}
\affil{Dept. of Physics, 
CCIS 4-183, 
University of Alberta, 
Edmonton, AB T6G 2E1, Canada; heinke@ualberta.ca}
\affil{Humboldt Fellow, 
Max Planck Institut f{\"u}r Radioastronomie, 
Auf dem H{\"u}gel 69, 53121 Bonn, Germany}

\author{F. Yusef-Zadeh}
\affil{Department of Physics and Astronomy and CIERA, 
Northwestern University, 
Evanston, IL 60208, USA; zadeh@northwestern.edu}

\begin{abstract} 

Variable emission from Sgr~A*, the luminous counterpart to the super-massive
black hole at the center of our Galaxy, arises from the innermost portions of
the accretion flow. Better characterization of the variability is important for
constraining models of the low-luminosity accretion mode powering Sgr~A*, and
could further our ability to use variable emission as a probe of the strong
gravitational potential in the vicinity of the
$4\times10^{6}\mathrm{M}_{\odot}$ black hole. We use the \textit{Herschel}
Spectral and Photometric Imaging Receiver (SPIRE) to monitor Sgr~A* at
wavelengths that are difficult or impossible to observe from the ground. We
find highly significant variations at 0.25, 0.35, and 0.5 mm, with temporal
structure that is highly correlated across these wavelengths. While the
variations correspond to $<$1\% changes in the total intensity in the
\textit{Herschel} beam containing Sgr~A*, comparison to independent,
simultaneous observations at 0.85 mm strongly supports the reality of the
variations. The lowest point in the light curves, $\sim$0.5 Jy below the
time-averaged flux density, places a lower bound on the emission of Sgr~A* at
0.25 mm, the first such constraint on the THz portion of the SED. The
  variability on few hour timescales in the SPIRE light curves is similar to
that seen in historical 1.3 mm data, where the longest time series is
available, but the distribution of variations in the sub-mm do not show a tail
of large-amplitude variations seen at 1.3 mm. Simultaneous X-ray photometry
from XMM-Newton shows no significant variation within our observing period,
which may explain the lack of very large variations if X-ray and submillimeter
flares are correlated.  \end{abstract}

\section{Introduction} 

Sagittarius A* (Sgr~A*) is the luminous source
\citep[L$\sim10^{-8}~\mathrm{L}_{\mathrm{Edd}}$;][]{Genzel10} associated with
the super-massive black hole at the center of our Galaxy
\citep[M=$4\times10^{6}~\mathrm{M}_{\odot}$, D=8.3 kpc;][]{Ghez08,Gillessen09}.
Due to its mass, relative proximity, and faintness, Sgr~A* is the premier target
for studies of strong gravity, low-luminosity accretion flows, and quiescent
    galactic nuclei. 

Variable emission from Sgr~A* arises from deep in the potential well of the
black hole in the innermost regions of the accretion flow
\citep{Baganoff01,Genzel03,Doeleman08,Fish11,Dexter14}. Thus, features in the
light curve of Sgr~A* could provide a powerful probe of both the physics of the
flow and the gravitational potential around the black hole, yet the nature of
the variability is not fully understood.

Constraining the radiative mechanisms responsible for the luminosity of Sgr~A*
is complicated by the difficulties associated with measuring the spectral
energy distribution (SED). At many wavelengths, Sgr~A* is either obscured by
the galaxy or confused with gas (radio and X-ray), dust (submillimeter), or
stars (near-infrared), and intrinsic variability imposes a need for simultaneous
observations in as many bands as possible.  Many groups have coordinated
multi-facility observing campaigns to constrain the shape of the
quiescent, or time-averaged, SED and the spectral shape of variable
emission \citep[e.g.,][]{Falcke98,Eckart04,An05,Eckart08,
Yusef-Zadeh06,Marrone08,Dodds-Eden09,Haubois12,Brinkerink15}. The quiescent SED rises from
centimeter to millimeter wavelengths, peaks around 0.8mm \citep[in flux density
units;][]{Marrone06a, Bower15}, and declines through the IR and X-ray ---the only
other wavelengths where  Sgr~A* has been clearly detected. The $S_{\nu} \sim \lambda^{-0.5}$
radio spectrum is consistent with optically thick, stratified synchrotron emission
\citep{deBruyn76}, and the increasing slope ($S_{\nu}\sim\lambda^{-1}$) near
the spectral peak, the ``submillimeter bump" \citep{Falcke98}, has been
interpreted as coming from the innermost regions of the accretion
flow \citep{Falcke98,Doeleman08,Dexter10}. The transition from optically thick
to thin emission appears to occur over a range of wavelengths in the
millimeter/submillimeter regime \citep{Bower15,Marrone06a}. 

Studies of the variability of Sgr~A* have revealed some patterns in the changes
between wavelengths. X-ray and IR monitoring has shown that X-ray flares are
accompanied by IR flares whenever there is simultaneous IR data
\citep{Hornstein07} but that IR flares are not always accompanied by X-ray
flares.  The relationship between millimeter/submillimeter light curves and
features in NIR/X-ray light curves is less well understood. Some report
evidence for increased emission in the millimeter/submillimeter after spikes in
the NIR/X-ray \citep{Yusef-Zadeh06,Marrone08,Eckart08}. These authors argue
that the delay is due to the adiabatic expansion of a synchrotron-emitting
plasma, whose peak emission shifts toward longer wavelengths as the expanding
blob cools and becomes less dense. Models including multiple expanding
synchrotron-emitting blobs have been tuned to provide adequate fits to
simultaneous submillimeter, NIR, and X-ray flares
\citep{Eckart06,Eckart09,Eckart12,Yusef-Zadeh06,Yusef-Zadeh08,Yusef-Zadeh09}.
These models often predict that the spectrum of observed flares should peak at wavelengths
$\lesssim$0.3 mm, impossible to constrain from the ground. The expanding blob
scenario is consistent with the results of \citet{Hornstein07}, who did not
observe a change in the NIR spectral slope during a flare. The absence of
a change in spectral slope can be explained with a non-radiative cooling
mechanism, such as adiabatic expansion \citep{Marrone08}. However, other groups
do report NIR spectral slope changes during flux increases
    \citep[e.g.,][]{Gillessen06}.

Other authors suggest that millimeter/submillimeter light curves are
anti-correlated with NIR/X-ray features \citep{Yusef-Zadeh10,Haubois12}. This
could be due to reduced millimeter/submillimeter emissivity caused by
a reduction of the magnetic field strength or a loss of electrons due to
acceleration or escape ---all of which are expected outcomes of a magnetic
reconnection event \citep{Dodds-Eden10,Haubois12}. Alternatively, the reduced
millimeter/submillimeter flux density coincident with NIR/X-ray features could
be due to obscuration of the quiescent emission region by the excited NIR/X-ray
emission region \citep{Yusef-Zadeh10}.  \citet{Dexter13} model time-dependent
emission from Sgr~A* and show that NIR/X-ray features and submillimeter
features arise from different electrons so are not necessarily related, yet
they demonstrate how cross-correlation analysis can produce spurious peaks.
Thus, not all reported correlations between IR and millimeter/submillimeter
wavelengths may be evidence for a physical connection.

Another challenge for ground-based studies of Sgr~A* variability is adequately
sampling the relevant timescales. In the NIR, a break in the power spectrum of
variations has been reported on timescales $\sim3$~h \citep{Meyer09}, while at
millimeter/submillimeter wavelengths there appears to be a characteristic
timescale for variations similar to the $\sim6$~h observing windows available
to Northern hemisphere submillimeter telescopes. Space-based observatories can
observe Sgr~A* for longer intervals and can provide more accurate and precise
probes of these important timescales \citep[e.g.,][]{Hora2014}.

Relatively little is known about Sgr~A* at the wavelengths probed by the
\textit{Herschel} Spectral and Photometric Imaging Receiver
\citep[SPIRE,][]{Griffin10}.  SPIRE observes in three bands simultaneously:
0.5 mm, 0.35 mm, and 0.25 mm. A few ground-based observations at
0.45 mm and 0.35 mm have been made when excellent weather provided adequate
  atmospheric transparency. At 0.45 mm, single dish measurements have detected
Sgr~A* at $\sim1.2$ Jy and at $\sim4$ Jy, although $\sim1$ Jy uncertainty in
the absolute flux density is incurred due to confusion with extended dust
emission \citep[e.g.,][]{Pierce-Price00,Marrone08,Yusef-Zadeh09}.
\citet{Marrone06a} made interferometric measurements at 0.43 mm that resolved
Sgr~A* from surrounding emission. Those measurements revealed a flat 1.3 mm --
0.43 mm spectral slope and detected variability of $\sim3$ Jy.  At 0.35 mm,
  atmospheric opacity and confusion with dust are even more severe, yet a small
number of measurements have been made from the ground that suggest variability
by a factor $\sim3$ \citep{Serabyn97,Marrone08,Yusef-Zadeh09}.

Both theoretical predictions from model-fits to multi-wavelength flare data
\citep{Eckart06,Eckart09}, and observational hints from sparse inhomogeneous
ground-based observations suggest that the variability of Sgr~A* in the SPIRE
bands may be stronger than the variability seen at $\sim$1.3 mm \citep[the
typical variability amplitude at 1.3 mm is $\sim$1 Jy on long
timescales][]{Dexter14}.  SPIRE provides a unique opportunity to test the model
predictions and to compile a uniform and sensitive dataset at 0.5 mm, 0.35 mm,
and 0.25 mm. In this paper we use 25.5 hours of \textit{Herschel} SPIRE data,
together with overlapping X-ray and 0.85 mm observations provided by
\textit{XMM-Newton}  and the CSO to monitor for variability and constrain the
spectral shape of flares.

\section{Observations and Reduction} 
The data we present in this paper were
collected as part of a multi-facility observing campaign to monitor Sgr~A*. The
participating observatories included \textit{Herschel}, CSO,
\textit{XMM-Newton}, and the SMA. 

\subsection{\textit{Herschel} SPIRE} \label{spiresec}
SPIRE data were collected in two $12.75$ hour blocks: the first from
2011 Aug 31 22:04 UT through 2011 Sep 01 10:51 UT; and the second from 2011 Sep
01 20:33 through 2011 Sep 02 9:20. Each interval includes 668 scans across the
   Galactic Center. Table \ref{obsTable} shows the observation identifiers
(ObsIDs) downloaded from the \textit{Herschel} Science Archive for this work.

\begin{deluxetable}{cc}
\tabletypesize{\scriptsize}
\tablecolumns{2}
\tablewidth{0pt}
\tablecaption{\textit{Herschel} SPIRE ObsIDs}
\tablehead{\colhead{First Interval}&\colhead{Second Interval}}
\startdata
1342227655&1342227733\\
1342227656&1342227734\\
1342227657&1342227735\\
1342227658&1342227736\\
\enddata
\tablecomments{Observations were split into two intervals separated by 1 day.
We reduced each interval individually due to the large computer memory demands
of the calibration algorithms included in HIPE.\label{obsTable}}
\end{deluxetable}

We reprocessed the SPIRE data products using the \textit{Herschel} HIPE
software to include the extended flux density gain calibration data products in
version 3.1 of the HIPE calibration tree. This step normalizes the response of
each bolometer integrated over the beam area, rather than to the peak flux
density, which is more appropriate for fields with extended emission. We also
chose to include the scan turnarounds in our reprocessing and map making. This
option provides additional points on the sky where bolometers make overlapping
measurements, increasing the constraints on the calibration algorithms.

For each interval, we concatenated all the Level 1 scans from each SPIRE array for
each ObsID array into a single Level 1 context to feed to the HIPE
destriper\footnote{We used the destriper included with the unreleased
development version of HIPE 14.0.2035, which provided improved convergence}
(Schulz et al. in prep.). The destriper iteratively determines offsets for all
scans crossing the mapped region on the sky. Each scan consists of many
detector readouts. The iterations stop when the variances of the readouts
within the boundaries of the map pixels cannot be further improved. By running
the destriper in ``perScan"-mode, individual offsets---we used a 0-degree
polynomial--- are fitted for each scan of a given detector, compensating for
any long-term variations in the scans. We ran the destriper twice, using the
output diagnostic table and destriped scans as inputs for the destriper on the
second iteration.  This provided small improvements.

Although our relative detector calibration is optimized for extended sources,
we produced maps calibrated in units of Jy beam$^{-1}$. To do this, we binned
the scans into groups of 4, and made a single map for each bin, resulting in
a time resolution of 4.6 min per map. We assigned the same sky coordinates to
each pixel in each map, taking care to center the location of Sgr~A* in the
central pixel. 

Preliminary review of the maps revealed motion of the flux density distribution
with respect to the pixels. This motion is due to insufficiently reconstructed
pointing drifts of the telescope that result in inaccurate sky coordinates
associated with each bolometer readout. Uncorrected, these drifts limit the
precision with which we can calibrate the bolometers and extract light curves.

We solved for pointing offsets as a function of time by shifting each map to
best align with the first map produced for each observing interval. Total
drifts over the 12.75 hour observing intervals were $\sim2''$ and $\sim1''$ for
the observations starting 2011 Aug 31 and 2011 Sep 01, respectively. This is
consistent with pointing uncertainties given by \citet{SanchezPortal14}.

After solving for the best-fit shifts, we updated the coordinates of the SPIRE
scans in HIPE and re-ran the destriper and our mapping routine. We iterated the
whole process once, and the results showed that our shifts had converged.
A small residual drift, $\sim0.2''$, remains in the data.

\begin{figure*}
\epsscale{1.15}
\plotone{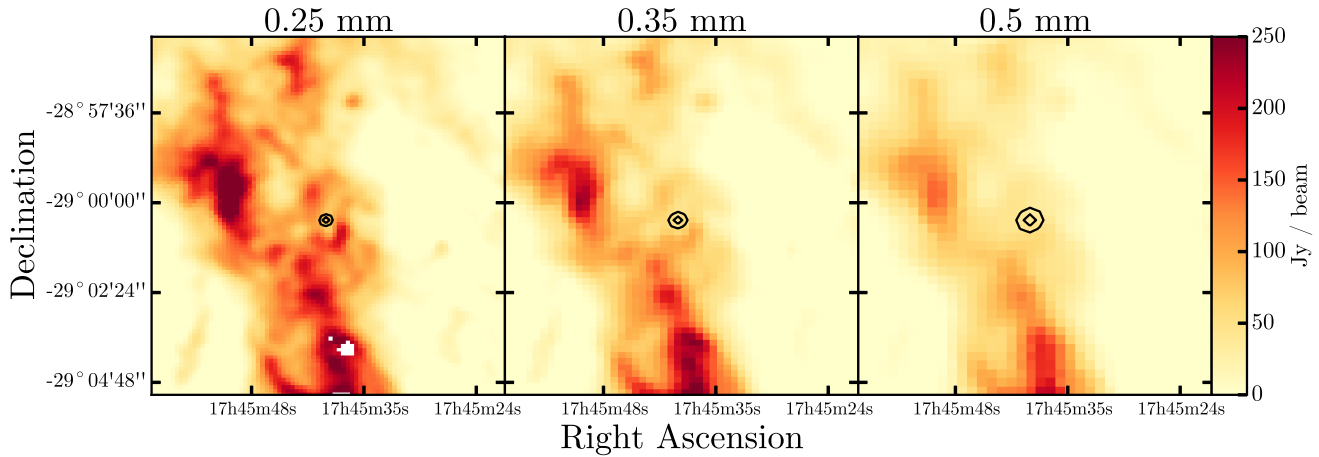}
\caption{\textit{Herschel} SPIRE maps of the galactic center. From left to
right are the maps at 0.25 mm, 0.35 mm, and 0.5 mm, respectively. Each map is
11$'$x10$'$. On each, we show the 80\% and 50\% contours of the
\textit{Herschel} beam centered at the location of Sgr~A*. Sgr~A* is not
resolved from its surroundings. For some map pixels the strong extended
emission from the Galactic Center region exceeded the dynamic range of the
SPIRE readout electronics in our chosen instrument setup, which was an accepted
trade-off to achieve a maximum sensitivity. This leads to some holes in the
0.25 mm map (white pixels)\label{maps}} 
\end{figure*}

Due to the bright extended emission of the Sgr~A complex at the Galactic
Center, and because of the relatively large beam size delivered by
\textit{Herschel} ($18''$, $25''$, and $36''$ at 0.25 mm, 0.35 mm, and 0.5 mm,
respectively) Sgr~A* is not separated from its surroundings. However, Sgr~A* is
expected to be the only intrinsically variable source in our maps \citep[there
is a magnetar, SGR J1745-2900, that is about $2.4''$ away from Sgr~A*, but the
magnetar was in its quiescent phase when these X-ray/submillimeter observations
were carried out;][]{Kennea13}. Therefore, we extract variability light curves
from difference images, subtracting the mean map of each 12.75 h observation
from the 4-min sub-maps.

We performed photometry on the difference maps by scaling a 1 Jy beam$^{-1}$
reference PSF to best fit our observations. We downloaded the reference beams
for each waveband from the SPIRE public
wiki\footnote{https://nhscsci.ipac.caltech.edu/sc/index.php/
Spire/PhotBeamProfileDataAndAnalysis}. Specifically, we combined our observed
difference maps ($D_{ij}$), a variance map created by calculating the variance
of all bolometer readouts contributing to a given map pixel
($\sigma^{2}_{ij}$), and the PSF ($P_{ij}$) as follows:

\begin{equation}
\bar{f}=\frac{\sum_{ij}(\frac{P_{ij}^2}{\sigma^{2}_{ij}})f_{ij}}{\sum_{ij}(\frac{P_{ij}^2}{\sigma^{2}_{ij}})},
\end{equation} 
which is the inverse variance weighted mean of the scale factor
\begin{equation} 
f_{ij}=\frac{D_{ij}}{P_{ij}}.  
\end{equation}
The variance of our measured scale factor can then be computed using
\begin{equation}\label{vareq}
\bar{\sigma}^{2} = \frac{\sum_{ij}(\frac{P_{ij}^2}{\sigma^{2}_{ij}})(f_{ij}-\bar{f})^{2}}{\sum_{ij}(\frac{P_{ij}^2}{\sigma^{2}_{ij}})}.
\end{equation}

We extracted light curves from the location of Sgr~A* and several reference
locations. Reference light curves should show no intrinsic variability so serve
as indicators of time variable systematic problems.  We chose reference
locations as follows: First, we generated 100 random locations within
2 arcminutes of Sgr~A*. From that set we excluded any points whose average flux
  density was not within a factor of two of the average flux density at the
position of Sgr~A*.  We also excluded points where the local spatial gradient
had a magnitude that was not within a factor of two of the gradient at the
location of Sgr~A*.  We then searched the remaining locations for a maximal set
with no two references within $40''$.  This last criterion ensures that the
0.5 mm beam does not overlap at the 50\% level for any of our reference
  locations.  Using this approach we identified 12 locations on the map for use
as references.

We noticed that many of our reference light curves were affected by a small
linear trend across the observing interval. This trend was largest in the 0.25
mm band where the average slope was measured to be $\sim-0.02$~Jy h$^{-1}$. The
steepest slope removed from our reference light curves was
$\sim-0.1$~Jy~h$^{-1}$. These values are consistent with the trends expected
given the residual pointing drift that remains in the maps and the flux density
gradients at the locations of our references. Since these drifts strongly
affect the appearance of the inter-band cross-correlations (resulting in
relatively high power over a large range of lags) we subtract a best-fit line
from each light curve. Unfortunately, this correction precludes a meaningful
test of whether our light curves are stationary \citep[a time series is
stationary if there are no changes in its mean value or variance and there are
no periodic components,][]{Chatfield89}. However, in Section \ref{discSec} we
quantify changes in the variance of our light curves in subintervals of
four-hour length.

Our calculated errorbars (Equation \ref{vareq}) were over estimated for each
location, including Sgr~A*. This was obvious given the magnitude of the point
to point variations in the light curves and the much larger size of the
calculated errorbars. The over-sized errorbars result from the way that our
variance maps are produced.  In HIPE, variance maps are produced by binning all
bolometer readouts that occur within a given pixel without respect to where
within a pixel a readout occurs. In regions of complex structure, such as Sgr
A, spatial flux density gradients will lead to variations in flux density
values within a pixel, inflating the variance.  To account for this, we scaled
the errorbars for each reference light curve to provide a good fit to
a constant zero-flux model (reduced $\chi^2=1$). Typical scale factors were
$\sim0.3$. We took the mean scale factor and applied it to the errorbars for
the light curve of Sgr~A*. This approach provides empirically accurate
errorbars that maintain appropriate relative size as a function of time and
location on the map.

\subsection{Caltech Submillimeter Observatory 0.85 mm}

Ground-based observations with the SHARC II camera at the CSO provided 0.85 mm
monitoring from 2011 Sep 1 04:25 UT through 09:04 UT, and from 2011 Sep 2 03:35
UT through 09:00 UT, overlapping each of our \textit{Herschel} observing
intervals. A $3'$ field surrounding Sgr~A* was observed with Lissajous scanning
of the telescope with an amplitude of  $100''$ and a period of  20~s
\citep{Yusef-Zadeh06,Yusef-Zadeh08}.  On both evenings, the conditions were
suitable for observation for the full periods, with clear skies or light
cirrus, low wind, and moderate humidity.  The zenith atmospheric opacity at 225
GHz was $\sim0.14$ on Sep. 1 and $\sim0.10$ on Sep. 2.  The telescope focus was
monitored and, as needed, adjusted during separate observations of point
sources, accounting for the gaps in the light curves; the larger gap around
6:00 UT on Sep. 2 was due to a brief observation of Sgr~A* at 0.35 mm which did
not yield useful results.

Data analysis, including absolute calibration, followed the method
described by \citet{Yusef-Zadeh09}.  Sgr~A* is not well resolved from
surrounding dust emission with the $19''$ resolution of CSO at 0.85 mm. This adds
$\sim1$ Jy uncertainty to the absolute flux level of Sgr~A* measured at 0.85 mm, but
the measurement of variations is much more precise.  Uncertainties for each
0.85 mm measurement were derived from the rms in the image, from which the mean
image and a gaussian at the position of Sgr~A* have been subtracted.

\subsection{\textit{XMM-Newton}} 

\textit{XMM-Newton} data were collected in two blocks, the first (ObsID
0658600101) from 2011 Aug 31 at 23:37 UT to Sep 1 at 12:58 UT, and the second
            (ObsID 0658600201) from 2011 Sep 1 at 20:26 to Sep 2 at 10:42 UT.
Sgr~A* was placed at the center of the \textit{XMM-Newton}/EPIC field of view
(away from any chip gaps). The medium filter, and full-frame mode, were used for
all three EPIC instruments. The more sensitive pn camera \citep{Struder01} had
exposures of 41.9 and 45.2 ks in the two observations, respectively.  The less
sensitive MOS1 and MOS2 cameras \citep{Turner01} had exposures of
48.6 and 52.3 ks in the two observations.  Below we focus on results from the
   pn camera; the MOS results were similar.

The data were processed with the XMM Science Analysis Software (version 11.0.0)
to select PATTERN $\leq$ 12, energies between 2 and 10 keV, and FLAG=0.  We
extracted light curves at 300 second binning from a radius of 10 arcseconds
\citep[as typical for Sgr~A*, e.g.,][]{Porquet03}, centered at the location of
Sgr~A*. This radius only encloses 50\% of the emission from Sgr A*
\citep{Read11}, yet includes a significant amount of contamination from
unrelated sources, both diffuse and point-like. In fact, the quiescent flux of
Sgr~A* (2-10~keV $L_X\sim2.4\times10^{33}$ erg s$^{-1}$) is only $\sim$10\% of
the flux enclosed within $10''$ \citep{Baganoff03}.

No statistically significant (3$\sigma$) flares were observed in any of the
EPIC light curves, and the highest points in each lightcurve did not correspond
with the highest points in other light curves.  The most interesting possible
peak occurred at 4.85 hours into the first observation, reaching
0.153$\pm$0.023 counts s$^{-1}$, compared to an average rate of 0.10 counts
  s$^{-1}$.  We can thus set an upper limit on the background
subtracted Sgr~A* flare luminosity during our observations (for flares of 300
seconds in length) of 7.6 times the quiescent value, or $L_X$ (2-10 keV)
$<1.8\times10^{34}$ erg s$^{-1}$; longer flares have stricter upper limits
($<6\times10^{33}$ erg s$^{-1}$ on average for 1 ks flares). We assume an
absorbed power-law spectrum for the X-ray flares, as seen for the quiescent
Sgr~A* spectrum \citep{Baganoff03}, with photon index of 2.7 and
$\mathrm{N}_{\mathrm{H}}=9.8\times10^{22}$~cm$^{-2}$.  Fits to the spectra of
X-ray flares from Sgr A* span a range of spectral indices, from 1.7 to 3.2;
changing the assumed spectral index in this range of photon indices alters the
upper limits by 10\% up or down.

\subsection{SMA 1.3 mm} 
In an attempt to provide overlapping 1.3 mm data, Sgr~A* was also observed with
the SMA. Unfortunately, it was afternoon in Hawaii during our \textit{Herschel}
and \textit{XMM-Newton} observations. SMA observing conditions are typically
worst in the afternoon  because the unstable atmosphere corrupts the
interferometer phases. Given the poor quality of the data we can only put
a $\sim$ 30\% upper limit on the amplitude of variations of Sgr~A* during our
observations. This corresponds to $\sim1$ Jy, which is about the size of the
largest variations seen at 1.3 mm \citep{Dexter14}.

\section{Results}\label{resultsSec}
We show our average SPIRE maps of the Galactic Center in Figure \ref{maps}. On
each map, we overlay contours of the \textit{Herschel} beam at the location of
Sgr~A*. While the beam size at 0.25 mm is smaller than at the longer
wavelengths, the dust emission at this wavelength is significantly stronger.
The net result is a more challenging measurement at 0.25 mm.  For an analysis
of the dust properties at the Galactic Center, using SPIRE maps as well as
additional far infrared data, see \citet{Etxaluze11}.

\begin{figure*}
\plotone{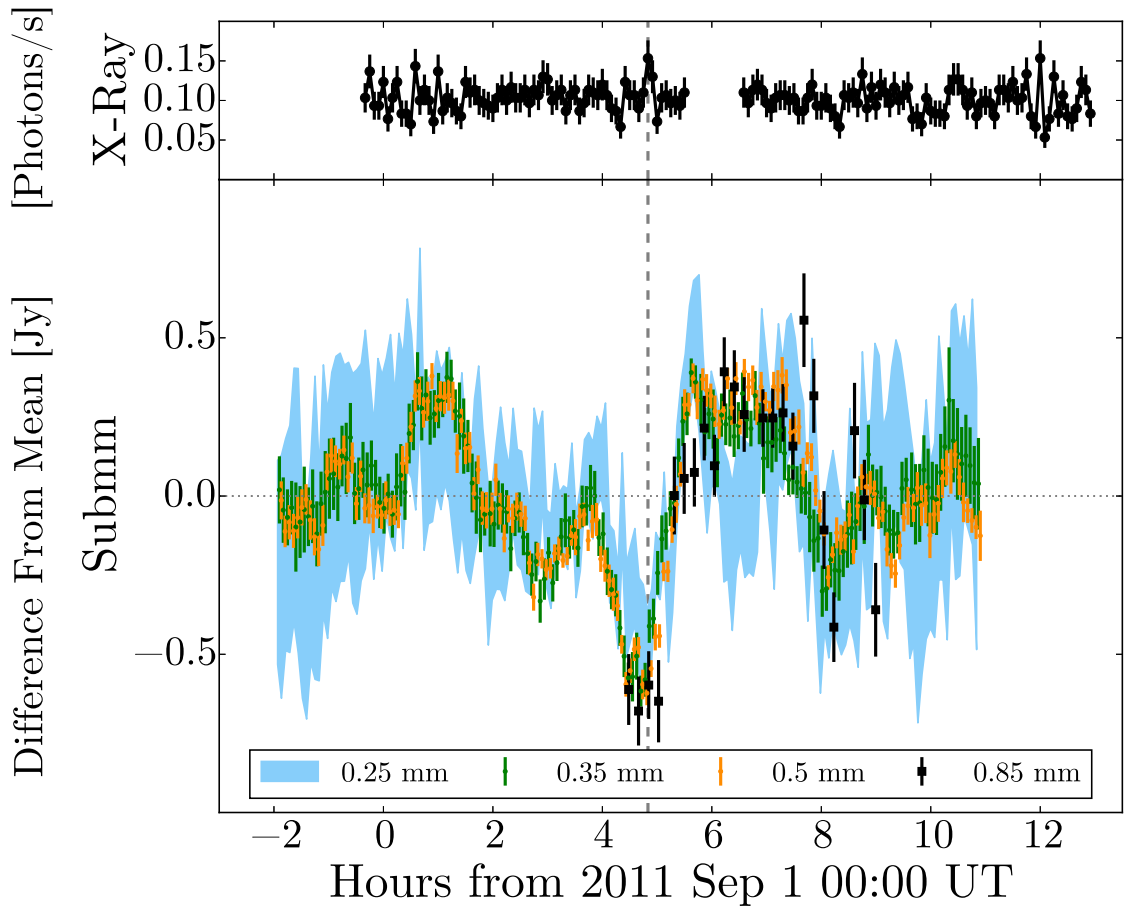}
\caption{Light curves from our first observing interval. Upper panel:
\textit{XMM-Newton} pn camera X-ray light curve. Lower panel: SPIRE and CSO
submillimeter light curves. For clarity in presenting four overlapping light curves, we
have employed two different plotting methods. The 0.25 mm light curve is shown
with a light-blue swath that indicates the 1-$\sigma$ confidence region. The
0.35 mm, 0.5 mm, and 0.85 mm light curves are displayed with dots and errorbars
indicating 1-$\sigma$ confidence. The SPIRE bands have each been offset slightly
in time to avoid overlap.\label{LC1}} 
\end{figure*}

\begin{figure*}
\plotone{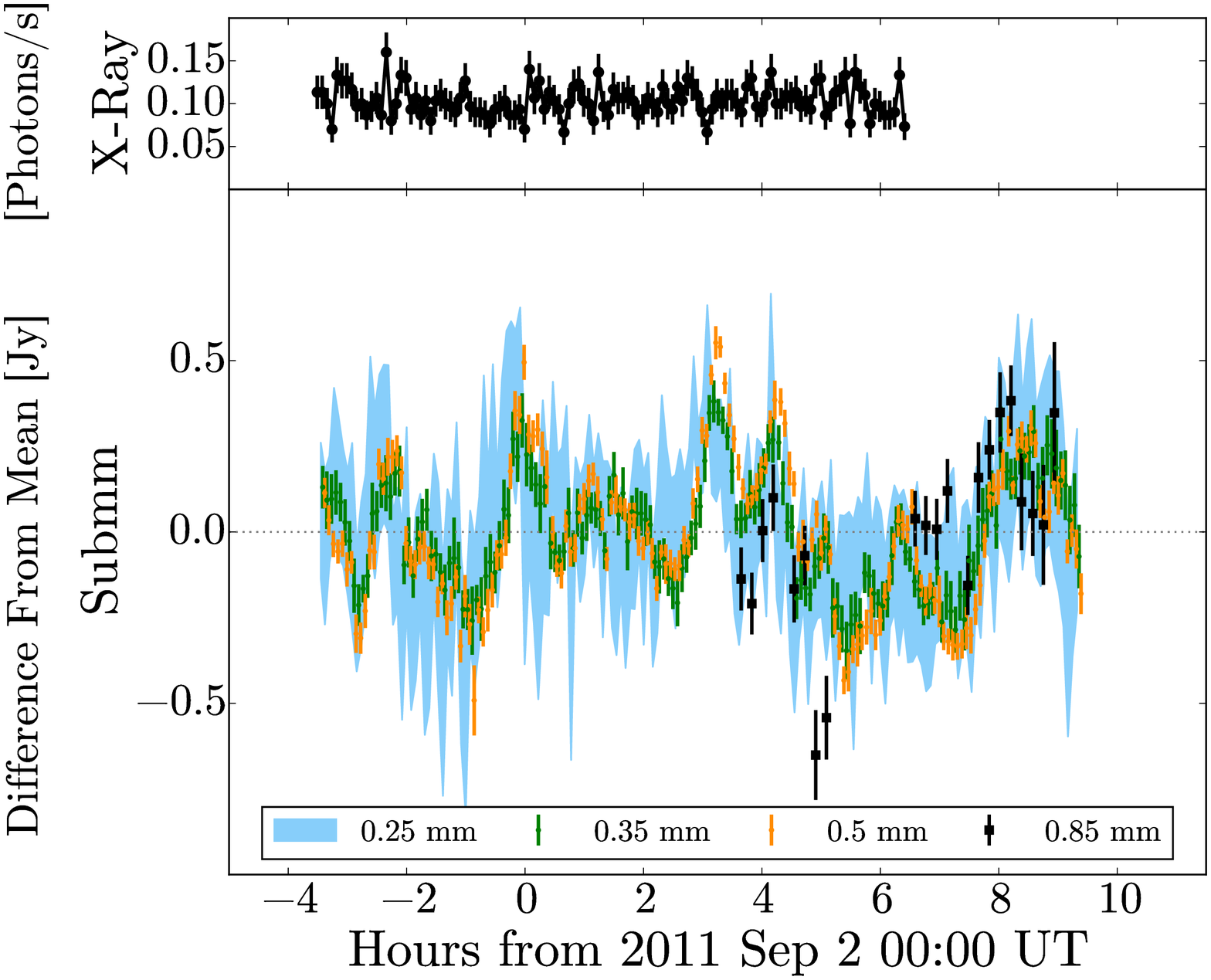}
\caption{Same as Figure \ref{LC1} but for our second observing interval.\label{LC2}} 
\end{figure*}

We show our Sgr~A* X-ray and submillimeter light curves for both observation
intervals in Figures \ref{LC1} and \ref{LC2}. There are significant variations
in all of the SPIRE bands. Ground-based 0.85 mm data closely track the SPIRE
bands during the first interval. The most significant feature, a flux density
decrement, occurs just before 05:00 UT on September 1st, and is captured by
both \textit{Herschel} and the CSO. The magnitude of the dip, $\sim0.5$ Jy, was
similar in all bands, and for the \textit{Herschel} bands corresponds to 0.6\%,
0.8\%, and 0.5\% of the flux density in the beam containing Sgr~A*. The dip
  coincides with a marginal feature in the X-ray light curve, which we
highlight with a vertical dashed line. This behavior is reminiscent of the
data reported in other studies that show $\sim0.6$ to $1$ Jy decrements in
millimeter light-curves correlated with flares in the near-infrared and
X-ray \citep{Yusef-Zadeh10, Dodds-Eden10, Haubois12}. 

\begin{figure*}
\plottwo{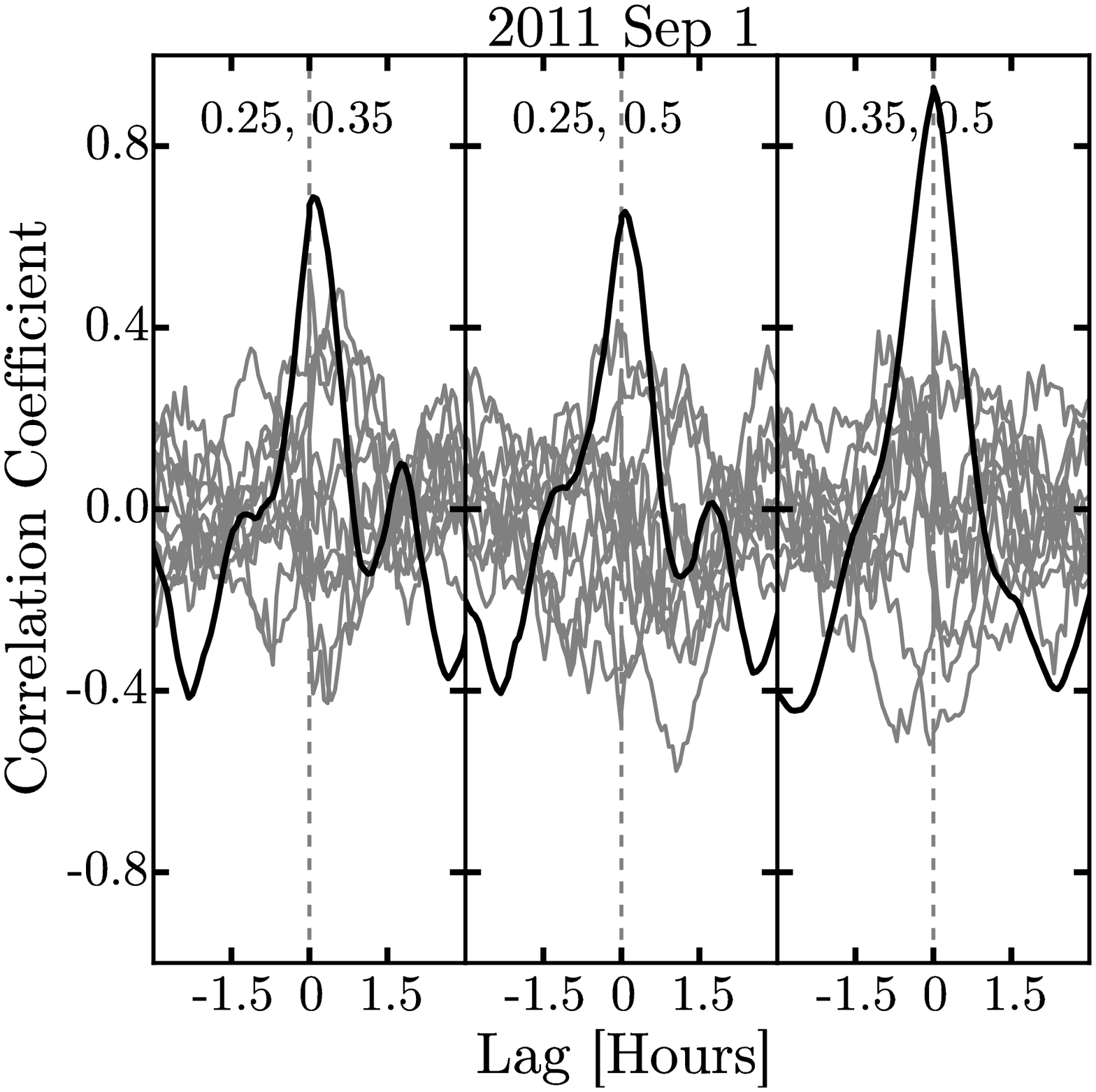}{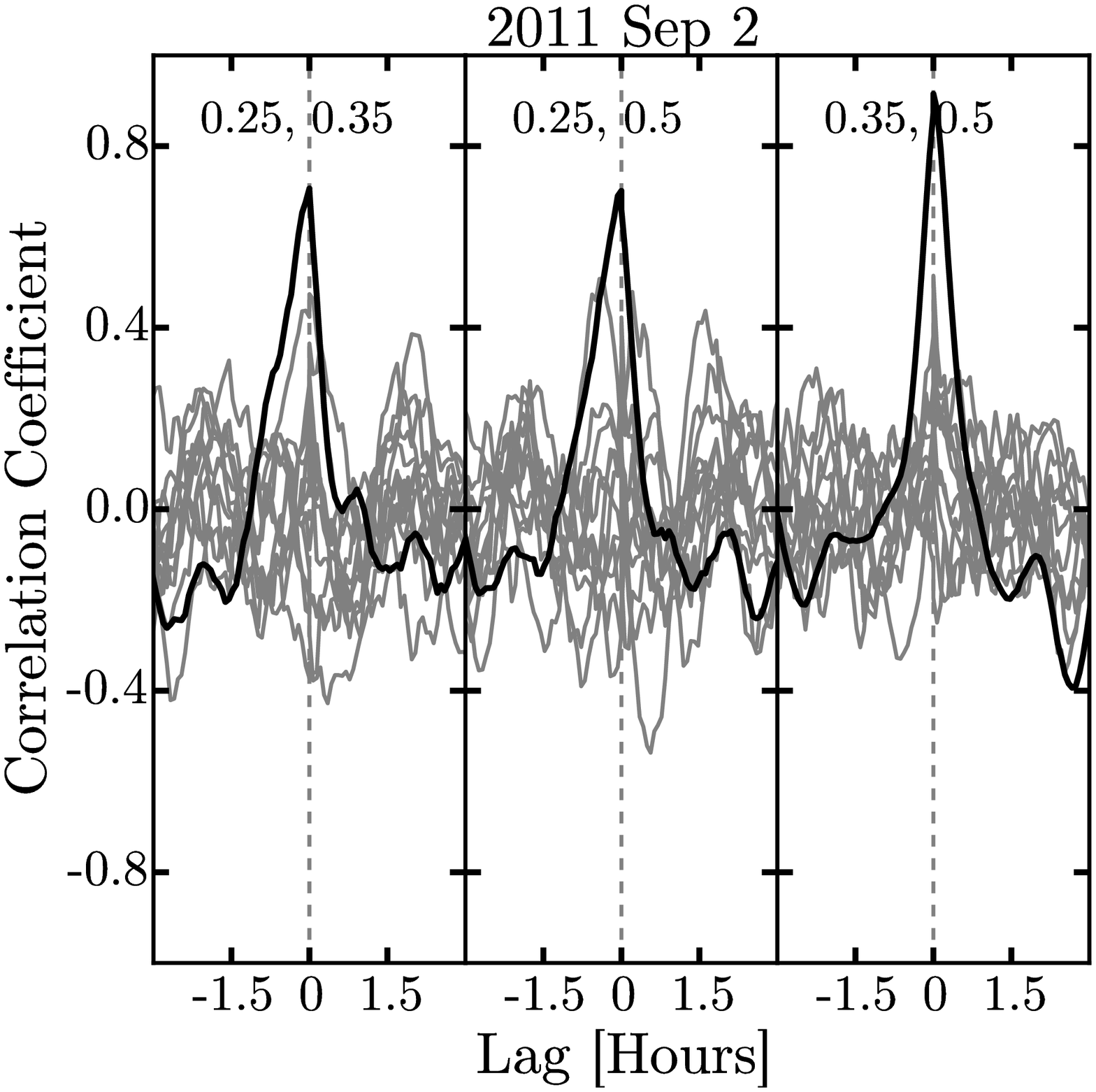}
\caption{Cross correlations for all three pairs of SPIRE bands for Sgr~A*
(black line) and reference locations (gray lines). Left: data from the first
observing interval. Right: data from the second observing interval. In each
panel, the left column shows the cross-correlation of the 0.25 mm light curve
with the 0.35 mm light curve, the middle column shows the cross-correlation of
the 0.25 mm light curve with the 0.5 mm light curve, and the right column shows
the cross-correlation of the 0.35 mm light curve and the 0.5 mm light
curve.\label{corr}} 
\end{figure*}

The significance of features seen in the SPIRE light curves is supported by
cross-correlation. In Figure \ref{corr} we show the cross-correlation of the
light curves for each pair of SPIRE bands, for each observing interval. We show
cross-correlations for Sgr~A* (black curves), and each of the 12 reference
locations (gray curves).

All pairs of Sgr~A* light curves are more correlated than pairs from the
reference locations.  This implies the presence of a shared signal, stronger
than the residual systematics that could result in spurious zero-lag
correlations for the references (e.g., pointing inaccuracies and thermal
drifts). The absence of dominant systematics in these $\sim0.5\%$ difference
measurements is also indicated by the agreement with the independent
measurements made by the CSO.  Cross-correlation peaks for curves including the
0.25 mm light curve from our first observing interval occur lagged by 4 min, or
  one sample. All the other cross-correlation curves show zero lag.  

\section{Discussion} \label{discSec}
\subsection{Variability amplitude compared to 1.3 mm}\label{varsec}
We observe strong variations in all three SPIRE bands with similar amplitude in
each.  We do not know the absolute flux density of Sgr~A* at any of these
wavelengths due to confusion with the surrounding dust emission (though
interferometer measurements at 0.43 mm in \citet{Marrone06a} show a minimum
flux density of 2 Jy in 4 epochs).  However, the negative deviation around 5 UT
in the first interval implies that for all three bands there must be a minimum
time-averaged flux density of at least 0.5 Jy, even at 0.25 mm where the SED is
not otherwise constrained. 

We now attempt to provide an empirical comparison of Sgr~A* variability at
SPIRE wavelengths to variability at 1.3 mm. \citet{Dexter14} provide
a detailed analysis of Sgr~A* variability at 1.3 mm, 0.8 mm, and 0.43 mm. They
demonstrated consistent variability amplitude characteristics between the
bands, though the 0.8 mm and especially the 0.43 mm characteristics were poorly
constrained due to the smaller number of measurements at those wavelengths. To
compare our light curves to 1.3 mm observations composed of many shorter
intervals of irregularly sampled data, we devised the approach described below.

\begin{figure}
\plotone{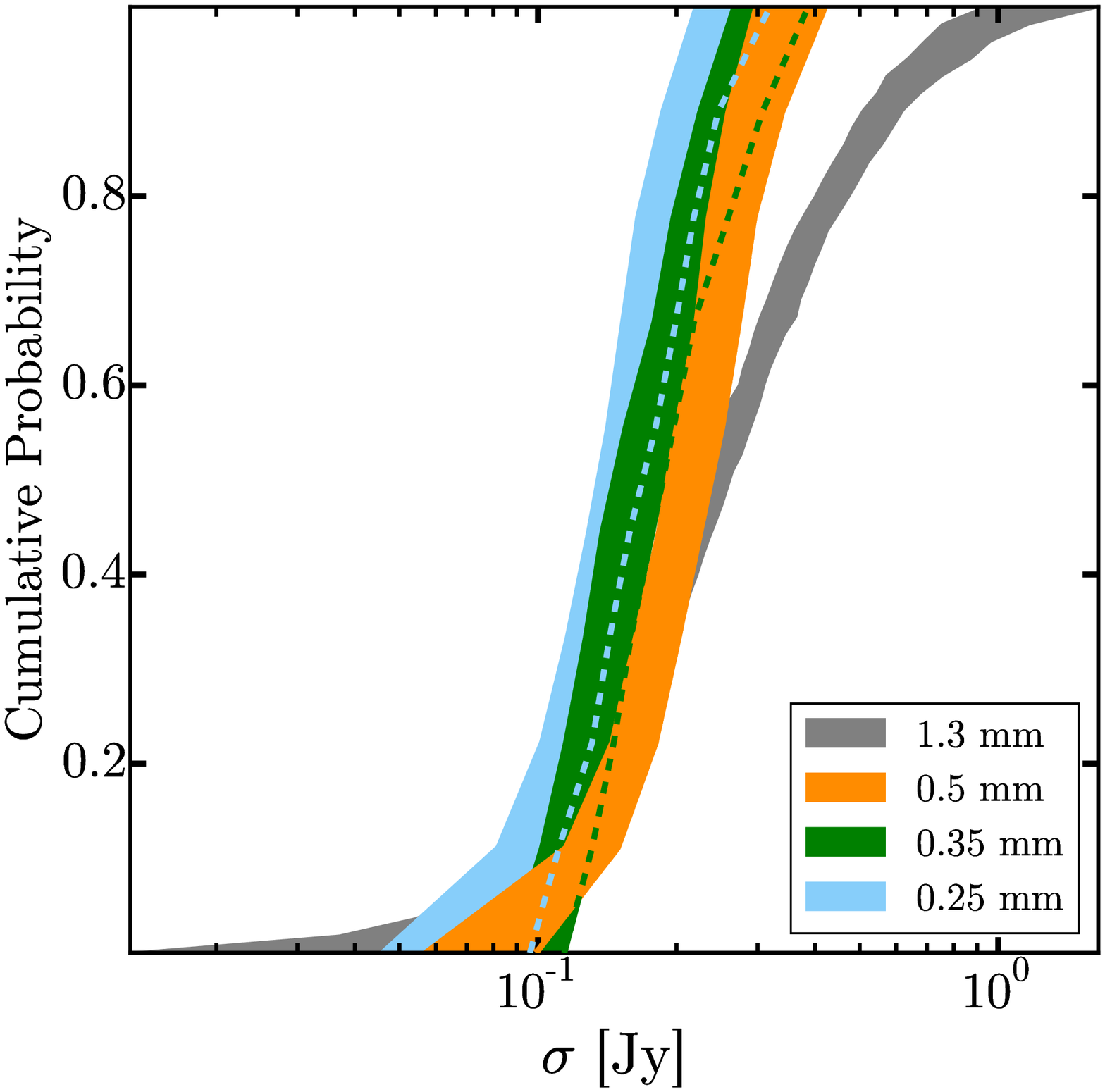}
\caption{Cumulative distribution of variability amplitude in overlapping 4-hour
blocks. The width of each swath contains the central 68\% of CDFs for each
light curve, generated by sampling the likelihood function of $\sigma$ for each
4-hour block. (See text for details.) Where obscured, dashed lines indicate the
edge of each swath.\label{cdfs}} \end{figure}

Using the SPIRE light curves, as well as the $\sim$70 h of 1.3~mm light curves
compiled by \citet{Dexter14}, we determine most-likely variability amplitudes
for overlapping 4-hour subsets of the data at each wavelength and compile them
into a distribution function.  The segmentation time is chosen to span typical
variations in the light curves, though our results are not very sensitive to
the choice.  We used overlapping segments, whose start times were separated by
at least 1.33 hours (66\% overlap), to provide additional measurements of
$\sigma$. Explicitly, for each four-hour block, we binned the data to 20-minute
time resolution for better signal-to-noise, subtracted the bin mean, and then
constructed the function 
\begin{equation} 
P_{\mathrm{block}}(\sigma | f_{i},\delta_{i}) = 
\Pi_{i} (\frac{1}{\sqrt{2\pi} (\sigma^{2}+\delta_{i}^2)})
e^{(\frac{-f_{i}^{2}}{2(\sigma^{2}+\delta_{i}^2)})}, 
\end{equation} 
for the probability that $\sigma$ is the typical variability amplitude
within the block, given the binned mean-subtracted flux density measurements
($f_{i}$) and their uncertainties ($\delta_{i}$).  We evaluated each
$P_{\mathrm{block}}$ at $\sigma=0$ to 2~Jy using 0.001 Jy steps. This range and
step size is wide enough to capture the most probable value in each block and
to finely sample the function. By taking the most likely variability amplitude
from each block, we can create a single cumulative distribution function (CDF)
for each band. In order to illustrate the range of possible CDFs that are
consistent with our data, we numerically sample the probability function for
each block 100 times and create 100 additional CDFs. Figure \ref{cdfs} shows the
region containing the central 68 CDFs created in this way for each wavelength.

Numerical simulations predict constant or increasing fractional variability
with decreasing wavelength in the millimeter/submillimeter portion of the SED
\citep[e.g.,][]{Goldston05, Dexter13, Chan15}. While the allowed regions for
the SPIRE CDFs of the variability amplitude overlap in Figure \ref{cdfs}, the
0.25 and 0.5~mm regions are mostly distinct, and there is a clear trend of
decreasing variability amplitude with wavelength, suggestive of a falling SED
from 0.5 to 0.25~mm.

The distribution of 1.3~mm variability is notably different from the SPIRE
curves in Figure~\ref{cdfs}.  There is a long tail of high-amplitude variations
not seen in the submillimeter light curves.  This may be the result of catching
Sgr~A* during a quiet state, as past ground-based measurements of Sgr~A* in the
0.45 and 0.35~mm atmospheric windows differed by several Jy
\citep{Dent93,Serabyn97,Pierce-Price00,Yusef-Zadeh06,Marrone06a,Marrone08}.
A quiet state is also suggested by the \textit{XMM-Newton} light curves shown
in Figures \ref{LC1} and \ref{LC2}, which show no flare event with a 2-10 keV
luminosity greater than $1.8\times10^{34}$ erg s$^{-1}$. The absence of
a larger flare in our X-ray light curves is consistent with the flare rate
inferred from more than 800 hours of \textit{Chandra} monitoring of Sgr~A*
\citep[$\sim 1$ day$^{-1}$ above $10^{34}$ erg s$^{-1}$;][]{Neilsen13}. The
1.3 mm light curve has a duration nearly three times that of the SPIRE light
  curve, and also extends over many years and therefore includes more of the
long term fluctuations that can be expected from AGN variability.

\subsection{0.35 mm - 0.5 mm color changes}
\begin{figure*}
\plottwo{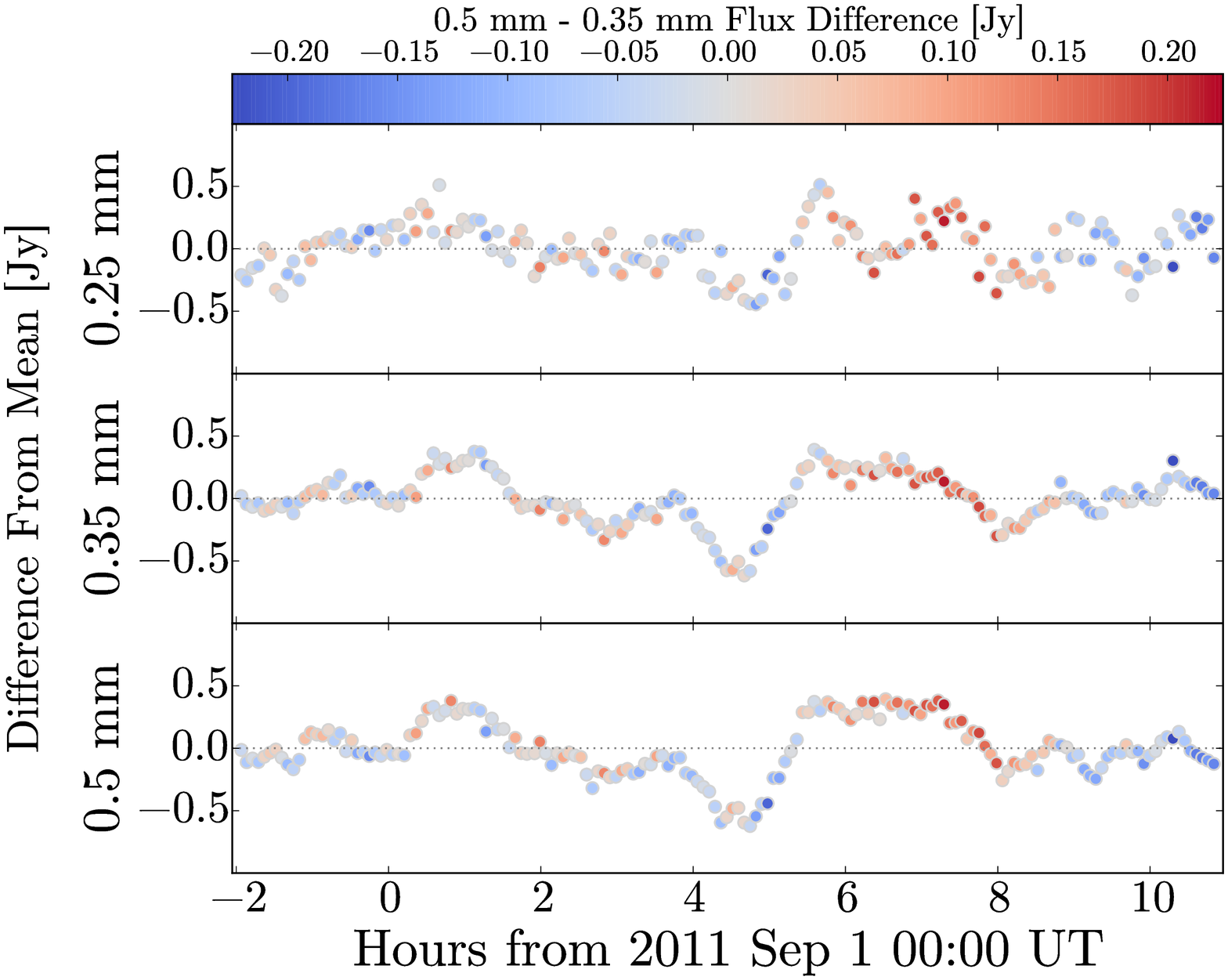}{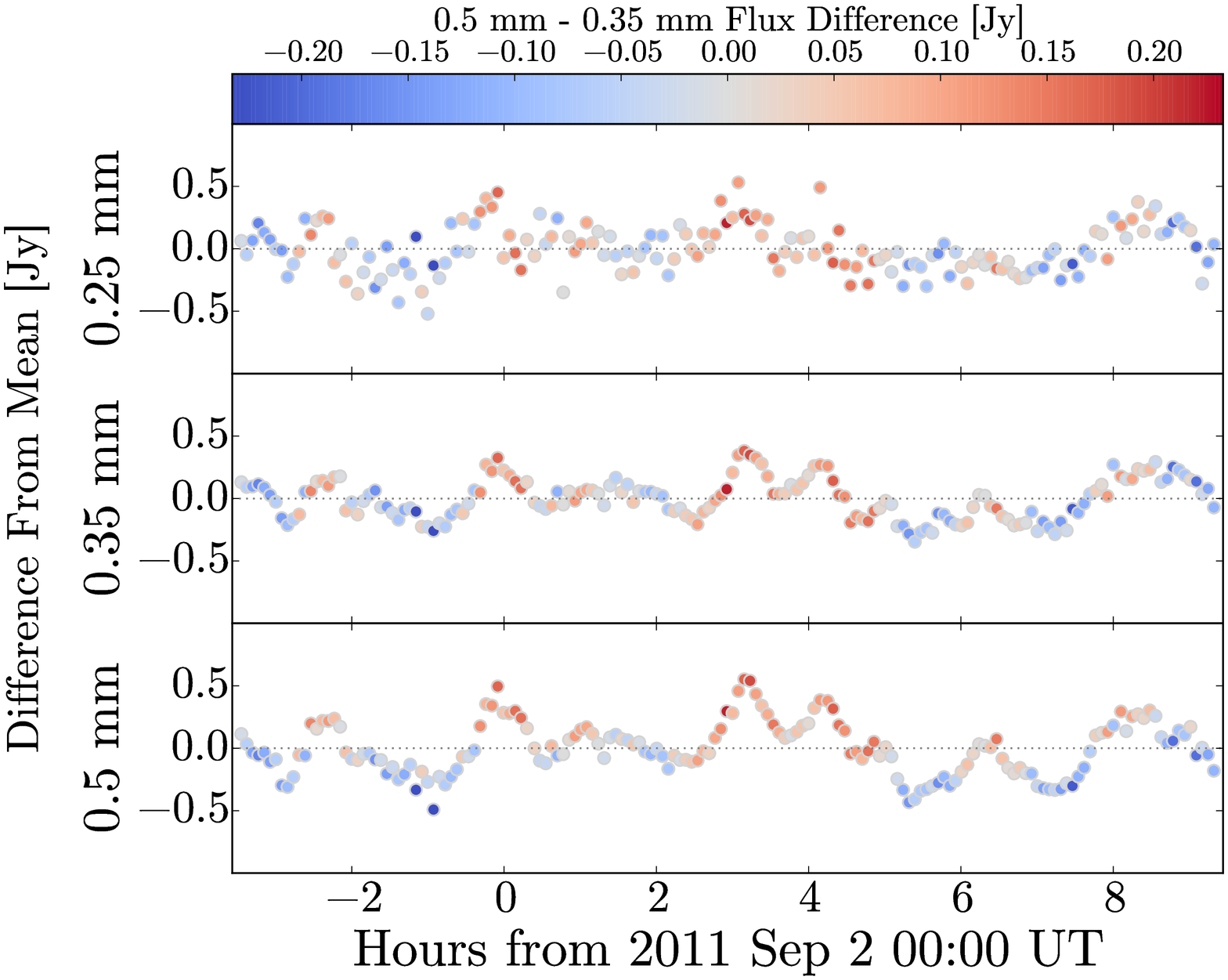}
\caption{SPIRE light curves with flux density measurements color-coded to show
the 0.5 mm - 0.35 mm flux density difference. The left panel shows the light
curves from the first interval and the right panel shows the light curves from
the second interval.  \label{colors}} 
\end{figure*}

In Figure~\ref{colors} we again plot the SPIRE light curves, but now show how
the 0.5--0.35~mm flux density difference changes with time. In the first
interval, we see a red color following our largest observed feature, from 05:00
to 08:00 UT.  Notably, the flare falls off even faster at 0.25~mm than it does
at 0.35 or 0.5~mm.  The reddened color as flux density decreases is suggestive
of a cooling process.  The expanding-blob model for features in the
submillimeter light curve of Sgr~A* should exhibit a blue-first fall off in
flux reminiscent of what we observe, yet the simultaneous rise at all
wavelengths is not consistent with the most naive blob models.  It is somewhat
harder to predict how color changes should manifest in more complex occultation
models where both absorption and adiabatic expansion take place simultaneously
\citep{Yusef-Zadeh10}. 

During our second observing interval, the light curves exhibit a clear pattern
of relatively red local maxima, and relatively blue local minima. This pattern
indicates a larger absolute amplitude of variation for the longer-wavelength
band. This pattern is less evident in the first interval, though there is some
hint of a similar pattern in the smaller flares and the largest flare is
brightest at 0.5 mm for most of its duration. This seeming change in the
spectrum of the flaring emission between the first and second intervals may
indicate that the quiescent 0.5--0.25~mm spectrum also changed, but without
a way to directly measure the absolute flux density of Sgr~A* we can only
speculate.

\subsection{Power spectrum analysis}
We analyzed the power spectra of variations in the SPIRE Sgr A*
light curves. To do this we combined the spectral averaging technique of
\citet[][]{Welch67} ---to provide a high fidelity estimate of the power
spectrum--- and the Monte Carlo fitting approach of \citet{Uttley2002} ---to
properly account for the effects of our sampling pattern, aliasing, and
red-noise leak on the shape of our power spectra.

\begin{figure}
\plotone{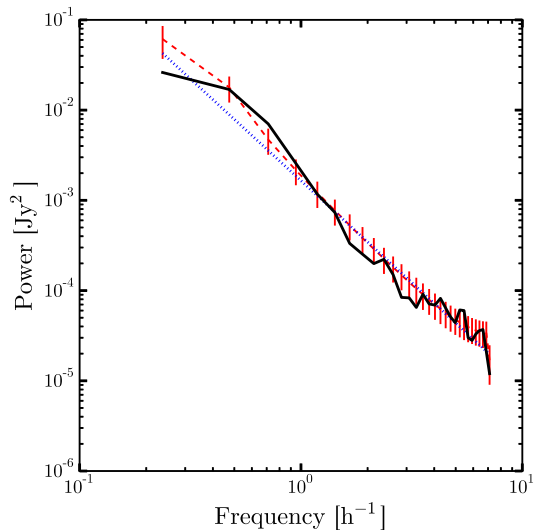}
\caption{Observed (black curve) and fitted power spectra for our 0.5~mm Sgr~A*
light curve. The power spectra for the 0.25~mm and 0.35~mm light curves are
similar, though less well determined. We used Welch's method to estimate the
power spectrum as discussed in the text. The red dashed curve and error bars
show the best fit spectrum, corresponding to $\beta=2.4$, resulting from our
Monte Carlo analysis based on \citet{Uttley2002}.  The green dotted curve
represents the best linear fit to the spectrum on log-log space, and suggests
$\beta=2.25$.\label{psd}} 
\end{figure}

Welch's method involves dividing a time series into overlapping
segments, apodizing and Fourier transforming each, and then averaging the power
at each frequency. We chose to use segment lengths 1/3 as long as our 12.75~h
observing intervals in order to closely match the time scale we used in our
analysis in Section \ref{varsec}.  We used 50\% overlap as suggested by
\citet{Press2002}. After applying Welch's method to each of our two
12.75~h intervals, we combined the results to produce one average power
   spectrum. Figure \ref{psd} shows the final power spectrum for our 0.5~mm
light curve. For clarity we do not show the 0.35~mm or 0.25~mm power spectra.
The are similar though as expected based on the shape of the light curves,
though less well determined.

We modeled our light curve as arising from a power law noise process with $P(\nu)
\propto \nu^{-\beta}$. To find the best-fitting power law slope, $\beta$, we
followed \citet{Uttley2002} and simulated a large number of light curves with
a given slope, sampled them according to the sampling pattern defined by our
observations, computed their power spectra in the same way as for our Sgr A*
light curves, and then used the distribution of simulated spectra to define
a goodness of fit metric.

For a given $\beta$, we used the method of \citet{Timmer95} to simulate light
curves. To ensure we captured the effects of aliasing and red noise leak in our
simulated spectra, we produced synthetic light curves that were 200 times as
long, sampled 10 times as frequently as our observed light curves. These were
then sampled at the same rate as our observations and divided into 100 pairs of
12.75 h light curves. From each light curve, a best fit-slope was subtracted,
   as this was a necessary step in the reduction of our observed light curves
(Section \ref{spiresec}).  Finally, a single power spectrum for each pair was
computed using the same approach used for our observed light curves. This
process is then repeated 100 times to yield $10^{4}$ simulated spectra.

\begin{figure}
\plotone{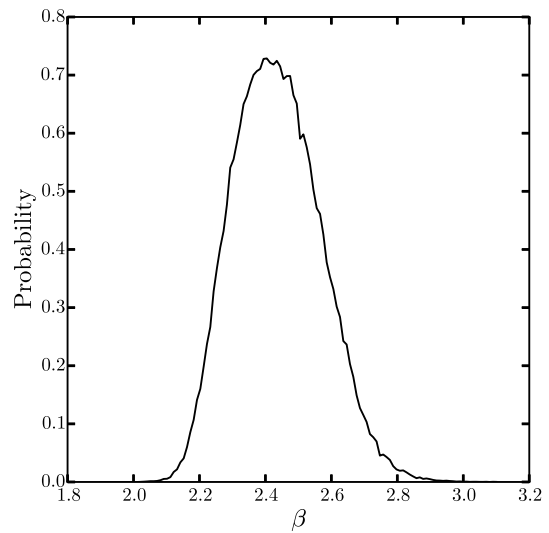}
\caption{Results of our Monte Carlo fitting procedure for the power-law index
of our observed power spectrum \citep[based on][]{Uttley2002}. Values below the
dotted horizontal line can berejected with greater than 95\% confidence.
\label{probplot}} 
\end{figure}

For each of the $10^4$ simulated spectra, we computed the quantity
\begin{equation}
\chi^{2}_{dist} = \sum\limits_{\nu}\frac{[P_{sim,i}(\nu)
- \overline{P_{sim}}(\nu)]^{2}}{\sigma_{sim}(\nu)^{2}},
\end{equation}
where $P_{sim,i}$ is a single simulated spectrum, $\overline{P_{sim}}(\nu)$ is the mean
of all the simulated spectra, and $\sigma_{sim}(\nu)$ is the standard deviation of
the spectra at each frequency \citep{Uttley2002}. We then computed a similar
quantity for our observed power spectrum, $P_{obs}(\nu)$,
\begin{equation}
\chi^{2}_{dist,obs} = \sum\limits_{\nu}\frac{[P_{obs}(\nu)
- \overline{P_{sim}}(\nu)]^{2}}{\sigma_{sim}(\nu)^{2}},
\end{equation}
scaling $\overline{P_{sim}}(\nu)$ and $\sigma_{sim}(\nu)$ by a common factor in
order to minimize $\chi^{2}_{dist,obs}$.  Then we compared
$\chi^{2}_{dist,obs}$ to the distribution of $\chi^{2}_{dist}$ defined by the
$10^{4}$ simulated spectra.  The rejection probability for a given $\beta$ is
taken as the percent of the simulated light curves with $\chi^{2}_{dist}$
smaller than $\chi^{2}_{dist,obs}$. We plot one minus the rejection probability
versus $\beta$ in Figure \ref{probplot}. Our best-fit power law slope is
$\beta=2.40$ with a 95\% confidence interval that spans from $\beta=2.16$ to
$\beta=2.73$. We show our best fit model spectrum and associated errorbars
in Figure \ref{psd}. For comparison with our Monte Carlo-based approach, we
also performed a basic fit of a line to our observed spectrum in log-log space.
In this case we recover $\beta=2.25$, which is less steep than we find
following \citet{Uttley2002}, yet still within the 68\% confidence interval.

Our derived spectral slope is very similar to the $\beta=2.3^{+0.8}_{-0.6}$
measured at 1.3~mm by \citet{Dexter14}.  Those authors also noted a break in
the power spectrum at 8$^{+3}_{-4}$~hours, which is a longer timescale than  we
have access to in our 12.75-hour intervals.  \citet{Meyer09} also found a slope
of $\beta=2.1\pm0.5$ at infrared wavelengths (mostly 2.2~$\mu$m), but with
a spectral break around 2.5~hours, and \citet{Hora2014} found consistent
characteristics at 4.5 $\mu$m. The consistency in slope from 1.3~mm, through
the SPIRE bands, out to the IR is not unexpected, as emission at all of these
wavelengths is expected to arise very close to the black hole, and therefore to
be subject to the same variations in the accretion process.

\section{Conclusions}

In this work we have presented the longest continuous submillimeter
observations of Sgr~A*, using 25.5~hours of data from the SPIRE instrument
aboard the \textit{Herschel Space Observatory}.  These data have provided
a first lower bound on the SED of Sgr~A* at 0.25~mm and characterized the
wavelength and temporal spectra of its submillimeter variations. While
\textit{Herschel} is no longer operational, the Atacama Large
Millimeter/Submillimeter Array (ALMA) can make ground-based measurements from
3 to 0.35~mm at high sensitivity, which can provide further constraints at
similar wavelengths.  In particular, the spatial resolution afforded by ALMA
will be adequate to isolate Sgr~A* from its surroundings, which was not
possible with \textit{Herschel}. Such data can more fully characterize the SED
of this source and its fractional variability. 

\acknowledgements
This work is based on observations made with Herschel, a European Space Agency
Cornerstone Mission with significant participation by NASA. We thank Chi-Kwan
Chan, Feryal {\"O}zel, and Dimitrios Psaltis for helpful discussions. DPM and
JMS acknowledge support from NSF award AST-1207752 and from NASA through award
OT1{\textunderscore}cdowell{\textunderscore}2 issued by JPL/Caltech. COH
acknowledges support from an NSERC Discovery Grant and an Alexander von
Humboldt Fellowship.

\clearpage
\end{document}